\documentclass[11pt]{article}

\usepackage[utf8]{inputenc}
\usepackage[T1]{fontenc}
\usepackage[margin=1in]{geometry}
\usepackage{times}
\usepackage{booktabs}
\usepackage{hyperref}
\usepackage{xcolor}
\usepackage{enumitem}
\usepackage[numbers,sort&compress]{natbib}
\usepackage{orcidlink}
\usepackage{microtype}

\hypersetup{
    colorlinks=true,
    linkcolor=blue,
    citecolor=blue,
    urlcolor=blue
}

\title{Visible, Trackable, Forkable:\\Opening the Process of Science}

\author{
    Sergey V. Samsonau\,\orcidlink{0000-0002-0835-2970}\\[6pt]
    Authentic Research Partners\\
    Princeton, NJ
}

\date{April 2026}

\begin{document}
\maketitle

% ============================================================================
% Abstract
% ============================================================================

\begin{abstract}
The way science is currently practiced shows conclusions but hides how they were reached. Researchers work privately, polish their results, publish a finished paper, and defend it. Errors are punished by retraction rather than corrected by amendment. Alternative directions are pursued through competing papers with no shared history. The reasoning, the dead ends, the trade-offs, the corrections: everything that would let others understand how a conclusion was reached is invisible. Two decades of open science reform have addressed this by opening specific artifacts: papers, data, code, notebooks, protocols. Each is valuable, but the unit remains a finished product. None opens the thinking process itself: the evolving sequence of questions, interpretations, dead ends, and direction changes that constitutes the actual scientific contribution.

This paper argues that opening the process of science (not just its outputs) would produce a step change in the speed of scientific progress, the accessibility of scientific reasoning, the trustworthiness of scientific claims, and the scalability of scientific quality assurance. We identify three properties the workflow needs: \textbf{visible} (the process is open, not just the product), \textbf{trackable} (every change is recorded and attributable), and \textbf{forkable} (anyone can branch from any point with shared history preserved). A visible, trackable flow is inherently verifiable: by humans, by automated tools, by AI agents. Software development adopted this flow decades ago, and the results (faster correction, broader contribution, maintained quality at scale) demonstrate the opportunity for science.
\end{abstract}


% ============================================================================
% 1. Introduction
% ============================================================================

\section{Introduction}
\label{sec:intro}

Two fundamentally different workflows exist for organizing complex intellectual work.

In the first, practitioners work privately for months or years, producing a finished artifact (comprehensive, polished, internally consistent). They submit it to gatekeepers who evaluate it. If accepted, it enters the permanent record, essentially unmodifiable. If an error is later found, the artifact is retracted: removed from the record, with the author's name attached to the failure. Others who wish to take the work in a different direction must start from scratch. Contribution is measured by the number of finished artifacts produced and how often others reference them.

In the second, work is continuously recorded. Every change, every decision, every correction is documented. Others can see what has been done, report problems, propose corrections, and receive credit for their contributions. If an error is made, it is fixed; the history shows both the error and the correction, and no one pretends the error did not happen. If someone disagrees with a decision made at step 47, they can take a complete copy of the work up to step 46 and pursue their alternative, with the full shared history preserved. Contribution is measured by what was actually done, not only by finished products shipped.

The first is how science is currently practiced. The second is how software development works. The difference is not a matter of tools or platforms. It is a difference in \emph{flow}: in the basic sequence of how work is done, how errors are handled, how contributions are tracked, and how disagreements are resolved.

This paper argues that science's workflow lacks three specific properties that software's workflow has, and that these missing properties (not peer review failures, not perverse incentives, not individual negligence) are the structural root of the problems science currently faces. The three properties are:

\begin{description}[style=nextline,leftmargin=2em,labelindent=0em,itemsep=0.5em,before=\medskip,after=\smallskip]
\item[\textbf{Visible.}] The process of research is open, not just the finished product. Currently, science publishes results but hides the work that produced them: the failed approaches, the corrected errors, the methodological trade-offs, the reasoning behind decisions. When people say ``open science,'' they mean open outputs. The process remains closed.
\item[\textbf{Trackable.}] Every change, every decision, every correction is recorded and attributable, at any granularity, from a single correction to a complete analysis. Currently, science records only one unit: the finished paper.
\item[\textbf{Forkable.}] Anyone can branch from any point in the work to pursue a different direction, with the shared history preserved. Currently, pursuing an alternative requires starting over, and the relationship between competing approaches is opaque.
\end{description}

A flow with these three properties is inherently verifiable. If the process is visible and every step is tracked, then anyone (a peer reviewer, an automated tool, an AI agent) can check the work at any point. Verification becomes a natural activity within the flow, not a separate gatekeeping event applied to a finished artifact months after publication. The key insight is that verification does not require a separate property; it requires visibility and trackability. You cannot check what you cannot see, and you cannot see what is not recorded.

These three properties are ordered by dependency. Visibility is the most fundamental: you cannot track or fork a process that is hidden. Trackability enables forkability (you can only branch meaningfully from a point you can identify in a recorded history). Together, they constitute a different flow, a different way of working that produces different outcomes.

\textbf{Related work.} A substantial literature has addressed opening specific artifacts of the research process. Open access made papers freely readable. The FAIR principles \cite{wilkinson2016fair} established standards for data findability, accessibility, interoperability, and reusability. Open Notebook Science proposed making the laboratory notebook (raw data, protocols, experimental records) publicly available in real time. The TOP Guidelines \cite{nosek2015top} formalized transparency standards for design, data, code, and materials. Munaf\`{o} et al.\ \cite{munafo2017} provided the most comprehensive catalogue of threats to reproducibility (HARKing, publication bias, p-hacking, low statistical power) and proposed reforms across methods, reporting, incentives, and evaluation, but each reform operates within the existing workflow: adding checkpoints, improving reporting, opening specific artifacts. Lyon \cite{lyon2016} identified transparency as a ``third dimension'' of open science beyond access and participation. Ram \cite{ram2013git} argued that version control systems like Git can improve reproducibility and transparency in research. Registered Reports pre-register methodology before data collection. Thibault et al.\ \cite{thibault2023} envision a ``truly collaborative research ecosystem'' and come closest to the present argument, noting that version control and forking could maintain transparent relationships between evolving research items; but their examples are protocols and code, still finished artifacts under version control rather than the thinking process itself.

These contributions share a common frame: they ask for more artifacts, or earlier artifacts, or more open artifacts, but the unit remains a finished product. Open access opens the paper. Open data opens the dataset. Open code opens the scripts. Registered Reports open the methodology before results are known. Even Open Notebook Science opens the laboratory record: what was done, what reagents were used, what measurements were taken. All of these are valuable. None of them open the thinking process itself.

The present paper argues for something different: opening the full thinking process. Scientists are not technicians following established procedures; their primary contribution is original thinking: forming questions, interpreting observations, constructing theories, designing methods, building original tools and instruments to study nature in new ways, making connections across domains, changing direction when evidence demands it. It is this evolving sequence of thought that constitutes the actual scientific contribution, and it is precisely this that the current workflow hides. Not just what data were collected, but what new observations arrived this week and how they changed the interpretation. Not just which code produced the results, but why an earlier approach was abandoned, what the intermediate results suggested, what new ideas emerged, what considerations led to a change in direction. Every week, all of it.

Scientists historically kept detailed notebooks that captured exactly this kind of evolving thought. Darwin's notebooks show decades of hesitation, contradiction, and gradual revision before the Origin of Species, not the clean logical progression textbooks present. Marie Curie's laboratory notebooks, still radioactive, preserved in lead-lined boxes at the Biblioth\`{e}que nationale, document daily measurements, procedural adjustments, equipment failures, and experimental dead ends alongside the discoveries. Einstein's working papers reveal eight years of wrong turns before general relativity. The clean narratives we associate with these scientists are retrospective reconstructions; the actual process was messy, and where the records survived, they became some of the most valuable documents in the history of science. Philosophers, historians, and sociologists of science have confirmed the pattern: published papers are retrospective reconstructions that do not preserve the temporal order or logic of the actual work.

That tradition of preserving the thinking process was largely lost when the terminal paper became the sole unit of scientific record. The three properties proposed here (visibility, trackability, and forkability) are structural properties of the workflow itself, not attributes of individual artifacts. The distinction is between opening outputs and opening the process that produced them.

% ============================================================================
% 2. How Science Currently Flows
% ============================================================================

\section{How Science Currently Flows}
\label{sec:current}

\subsection{The Terminal Artifact}
\label{sec:terminal}

The fundamental unit of science is the paper: a finished, polished document presenting a complete argument. The workflow that produces it is linear and terminal: conceive $\to$ investigate $\to$ analyze $\to$ write $\to$ submit $\to$ review $\to$ publish. Each step feeds the next, and the output of the final step (the published paper) is the only part of the process that enters the public record.

Everything before publication is invisible. The failed experiments, the abandoned approaches, the corrected calculations, the methodological trade-offs, the intermediate results that shaped the final analysis: all of this is discarded or buried in lab notebooks that no one outside the research group will ever see.

This was a reasonable design for a world in which publication was expensive, distribution was physical, and the pace of research was slow enough that a small number of expert reviewers could plausibly evaluate each manuscript. That world no longer exists: AI generates plausible manuscripts faster than reviewers can check them, research teams span continents, and public trust depends on a process no one can see.

\subsection{Error as Catastrophe}
\label{sec:error}

Because the paper is a terminal artifact (finished, published, defended), errors discovered after publication have no natural resolution path. The available options are extreme: retraction (removing the paper from the record entirely) or erratum (a brief correction that may not be noticed by subsequent citers). There is no mechanism for versioned correction, no standard way to indicate ``Section 3.2 has been updated; here is what changed and why.''

This creates a system in which error is catastrophic rather than routine. A bug in software is unremarkable; every software release has bugs, and the response is a patch. An error in a published paper can severely damage a career. The asymmetry is not because scientific errors are inherently more serious than software bugs; it is because the workflow provides no mechanism for graceful correction.

The consequences are predictable. When error is catastrophic, people hide errors. When hiding errors is rational, the system selects for the appearance of precision over actual precision. Researchers learn to perform rigor (citing papers they have not fully read, reporting analyses that confirm their expectations, presenting clean narratives that conceal the messy reality of investigation) because the alternative is risking catastrophe \cite{ioannidis2005,lipton2018troubling}.

The contradiction with what science teaches is stark. Popper's entire philosophy rests on the principle that scientific knowledge advances through conjectures and refutations; falsification is the engine of progress, not a failure to be punished. Every textbook teaches that errors are data. Then students look at actual practice: errors hidden, errors punished by retraction, clean narratives with no wrong turns. The lesson students actually learn is not the scientific method. It is that the appearance of rigor matters more than the reality.

The negative results problem is a direct consequence. Experiments that disprove a hypothesis, approaches that fail, results that are inconclusive: all invisible in the current flow because only finished papers with positive results get published. The problem was named ``the file drawer problem'' in 1979; as of 2025, a consensus paper calls all previous efforts to fix it ``fragmentary and largely unsuccessful'' \cite{curry2025}. Every attempted solution operates within the terminal artifact flow, creating a new category of publishable artifact rather than opening the process that would make negative results naturally visible as part of the record. In a visible flow, there is no need to ``publish'' negative results. They are simply part of the history. And what appeared to be a dead end at the time may later be reinterpreted as an important observation when new context emerges, but only if the record exists to be revisited.

\subsection{The Reproducibility Question}
\label{sec:reproducibility}

The ``reproducibility crisis,'' the finding that a large fraction of published results in psychology, biomedicine, economics, and other fields cannot be independently reproduced \cite{osc2015}, is real, but the way it is framed reveals the problem more clearly than the crisis itself.

The standard framing treats reproducibility as a test applied after publication: a different team attempts to obtain the same result using the description in the paper. When they cannot, the result is declared ``not reproducible,'' and the discussion turns to better methods, stricter statistics, pre-registration of hypotheses, and more careful reporting.

But what this test actually measures is revealing. A team is attempting to re-derive a result from a polished prose description of the process, a description that, by design, omits the intermediate steps, the parameter choices that were explored and discarded, the preprocessing decisions that shaped the data, and the dozens of small methodological choices that collectively determine the outcome. The paper is a \emph{narrative about} the process, not a \emph{record of} the process. Reproducing a result from a narrative is a fundamentally different task from following a tracked sequence of documented steps.

In a visible, trackable flow, ``reproducibility'' is not a separate heroic effort performed years later by a different team. It is a property of the workflow itself. If every step is recorded (every data processing decision, every parameter choice, every analytical branch), then reproducing the result means following the record. The same way a well-documented software build is reproducible by construction: you run the build, and it produces the same output, because every step is specified and tracked.

The reproducibility crisis, in this framing, is not evidence that science is doing bad work. It is evidence that the workflow makes good work unreproducible by hiding the process. The question ``can this result be reproduced?'' is the wrong question; or rather, it is a question that only needs to be asked when the process is invisible. In a visible flow, the question answers itself.

\subsection{Contribution as Authorship}
\label{sec:contribution}

Science measures contribution by authorship on published papers. This is a coarse metric that conflates many different kinds of work: the person who designed the experiment, the person who ran it, the person who analyzed the data, the person who wrote the paper, the person who provided funding, and the person who reviewed drafts are all listed as ``authors,'' with ordering conventions that vary by field and carry ambiguous signals.

More importantly, the authorship metric excludes anyone who does not produce a complete paper. A researcher who identifies an error in someone else's methodology and proposes a correction receives no credit. A student who spends six months on a project that produces useful negative results but no publication has nothing to show for it. A citizen scientist who contributes meaningful analysis to a research question has no standard way to document or verify that contribution.

In software, contribution is granular and visible. A developer's profile shows exactly what they contributed: which projects, which changes, when, and how those changes were evaluated. A teenager in Lagos who fixes a bug in a project used by millions has a verified, visible record of that contribution, identical in form to the record of a senior engineer at a major company.

\subsection{Disagreement as Competition}
\label{sec:disagreement}

When researchers disagree about methodology, interpretation, or direction, the current workflow offers one resolution mechanism: competing papers. Group A publishes a paper arguing for interpretation X; Group B publishes a paper arguing for interpretation Y. The relationship between the two lines of work is reconstructed by readers from citation patterns and prose: ``Smith et al.\ (2023) argue that\ldots, however, we contend that\ldots''

There is no shared record of where the two approaches diverged, what they share, or what specific assumption or decision separates them. A reader encountering both papers must do substantial detective work to understand what is actually in dispute versus what is common ground. The structure of the disagreement is invisible.

% ============================================================================
% 3. How Software Flows Differently
% ============================================================================

\section{How Software Flows Differently}
\label{sec:software}

\subsection{The Visible Flow}
\label{sec:visible}

The most basic difference between how software and science work is what others can see. When a software project is developed in the open (as the majority of the world's most important software now is), the process itself is public. Not just the finished product: the process. The discussions, the false starts, the disagreements, the bugs, the fixes, the reasoning behind every decision.

This is not transparency as a reporting obligation. It is transparency as a working method. The process is visible because the tools that enable the work (version control, issue trackers, code review systems) are inherently public by default. Visibility is not an addition to the workflow; it is the workflow.

The consequences for trust are direct. When someone asks ``why should I trust this software?'', the answer is not ``because a prestigious institution produced it.'' The answer is ``look at the history.'' The development record is the evidence. Anyone (a security researcher, a competitor, a curious student) can examine it. Trust is based on the visible record, not on the reputation of the producer.

Science inverts this. When people say ``open science,'' they mean the published paper is freely readable. The process that produced it remains as hidden as it was before the open-access movement. The output is open. The process is closed.

\subsection{The Trackable Flow}
\label{sec:trackable}

When a software developer changes a single line of code, the change is recorded with: what was changed (a precise diff showing the before and after), who changed it, when, and why (a commit message explaining the purpose). This is not optional. It is how the tool works. The full history of a project, from the first line of code to the current state, is available, searchable, and attributable.

This history is not just an audit trail. It is a working tool. When a bug is discovered, the history shows when the bug was introduced, by whom, and in what context. When a decision is questioned, the history shows why it was made and what alternatives were considered. When a new contributor joins the project, the history provides the context they need to understand the current state of the work.

Consider matplotlib, a widely-used Python visualization library. Its public repository contains tens of thousands of recorded changes spanning nearly two decades. An AI agent (or a human) can read this history and reconstruct the story of the project: how it started as one researcher's replacement for MATLAB's plotting capabilities, what design trade-offs were made as the user base grew, which recurring problems drove architectural changes, and how the project evolved from a single-author tool to a community-maintained standard with nearly two thousand contributors.

No equivalent record exists for any major scientific research program. The history of climate science, of CRISPR development, of the search for gravitational waves: these exist only as sequences of published papers, supplemented by informal accounts. The actual process (the daily work, the dead ends, the corrections, the decisions) is lost.

Because software's history is visible and tracked, it is also verifiable, not as a separate activity, but as a natural consequence of the flow. Automated checks run on every change: linters that flag suspicious patterns, type checkers that catch logical inconsistencies, test suites that verify expected behavior. These tools do not replace human judgment. They augment it by catching mechanical errors continuously, as they are introduced, rather than months later. Verification is what you \emph{do} with a visible, tracked record. It is not a separate property; it is the activity the flow enables.

\subsection{The Forkable Flow}
\label{sec:forkable}

When the maintainers of matplotlib decline a proposed change (say, a new color palette system), the proposer has options. They can argue their case in the public issue tracker. They can submit a modified proposal. Or they can fork: take a complete copy of the project and its entire history, make their change, and develop their version independently.

Forking is not hostile. It is a structural mechanism for resolving genuine disagreements. Both the original project and the fork share a common history up to the point of divergence. Anyone can compare them to see exactly where they differ and why. Sometimes forks are reabsorbed into the original project. Sometimes they become independent projects. Sometimes they demonstrate that the alternative approach was superior, and the original project adopts their changes. The structure makes all of this visible and navigable.

Now imagine this applied to a research project. A climate modeling group publishes their methodology and results as a trackable, versioned body of work. Another group disagrees with a specific assumption (say, the parameterization of cloud feedback). Instead of publishing a competing paper that re-derives everything from scratch, they fork: take the shared work up to the point of disagreement, substitute their alternative assumption, and develop the implications. The full history shows what is shared, where the divergence is, and what each approach produces.

This is not how science currently works, but it is how science \emph{could} work, and the result would be that the structure of scientific disagreement becomes transparent rather than opaque. Lakatos argued that science consists of competing research programmes that coexist, each with a shared hard core of assumptions and a protective belt of auxiliary hypotheses that evolve over time. Forking is the structural mechanism for what Lakatos described philosophically: the point where programmes diverge, the assumptions they share, and the choices that separate them would all be visible in the record. A forkable flow does not resolve disagreements; it clarifies them. The shared work is visible, the point of divergence is explicit, and the consequences of each choice are traceable. Clarity about where genuine disagreement lies is often more valuable than premature resolution.

% ============================================================================
% 4. What Changes with a Different Flow
% ============================================================================

\section{What Changes with a Different Flow}
\label{sec:changes}

\subsection{Trust Shifts from Authority to Evidence}
\label{sec:trust}

The current workflow communicates finality: a paper is either published (presumed correct) or retracted (presumed fraudulent). There is no status for ``corrected in Section 3'' or ``better methodology now exists.'' This binary forces a pretense of settled truth that science cannot deliver, and the public has noticed. During COVID-19, scientific guidance contradicted itself not because science was broken but because science was updating understanding as evidence accumulated. The workflow had no way to present this honestly. Each statement carried the authority of finality; when the next one contradicted the last, the public did not see science working. They saw it failing.

Philosophers of science have argued against this pretense for decades. Kuhn showed that science undergoes revolutionary paradigm shifts. Popper argued all knowledge is conjectural. Yet the workflow communicates the opposite.

A visible flow dissolves the pretense. The claim becomes not ``this is true'' but ``this is where the evidence currently points, and here is the full record showing how we got here.'' Trust shifts from ``trust the institution'' to ``examine the record.'' Anyone can assess whether the reasoning is sound, the corrections were honest, and the conclusions follow from the work.

\subsection{AI Makes the Visible Process Readable}
\label{sec:ai}

A visible process is only useful if someone can read it. A software repository with tens of thousands of commits is technically public, but practically inaccessible to most people. The volume is overwhelming, the notation is specialized, and the context required to interpret individual changes is substantial. For most of the history of open-source software, process visibility served primarily those who were already deep in the project.

Generative AI is beginning to change this. A large language model can already read the commit history of a software project and produce an intelligible narrative; code changes have a relatively constrained structure that current models handle well. Scientific reasoning involves more ambiguity, domain-specific judgment, and contextual interpretation, and the ability to synthesize entire research histories into plain-language narratives is still emerging. But the trajectory is clear, and the argument does not depend on AI reaching that capability today: a visible, trackable research record is valuable to human readers right now. AI readability is a bonus that strengthens the case over time.

Applied to science, the implications are significant. If a climate modeling project maintained its full research history in a visible, trackable format (every data collection decision, every model assumption, every parameter change, every correction), then anyone could ask an AI agent to explain the entire journey: which approaches were tried and why they were abandoned, what trade-offs were made, which pitfalls the researchers encountered, what limitations they acknowledged, where they were uncertain, and how the final conclusions emerged from this messy, honest, human process. Not a summary written by the authors. Not an interpretation filtered through a journalist. The actual record, examined end to end, explained in plain language.

This makes the intermediary optional. Currently, public understanding of science depends on journalists who summarize, experts who interpret, institutions that vouch. In a visible flow augmented by AI, a citizen can examine the research record directly: the reasoning, the choices, the limitations, the uncertainties. This is not a utopian fantasy. It is exactly what already happens with open-source software: when a security researcher questions whether a library is safe, they examine the commit history, the issue tracker, the full process. AI tools now make this examination faster and more accessible. The same capability, applied to science, would shift public trust from conclusion-based to process-based.

\subsection{Error Becomes Routine, Contribution Becomes Granular}
\label{sec:error-routine}

In a visible, trackable flow, finding and fixing an error is a normal part of the process, the equivalent of a commit that says ``fixed off-by-one error in Table 3 sample sizes.'' The error and its correction are both visible. No retraction, no catastrophe, no career damage. The history shows what happened and how it was resolved. The moral framing (error as sin) dissolves, because the workflow treats error the way engineering treats bugs: as expected, detectable, and correctable.

And every contribution is recorded: the original research, the correction of an error, the review of a proposed change, the addition of a new analysis, the identification of an inconsistency. Credit is proportional to actual work, not to position on an author list. A retired chemist who reviews methodology, an undergraduate who replicates an analysis, an independent researcher who identifies an error and proposes a correction: each contribution is verified, visible, and credited, because the flow records it automatically.

\subsection{Metrics Dissolve and Reform}
\label{sec:metrics}

The h-index, impact factor, and citation count are built on two units: the paper and the citation. Both dissolve in a visible, trackable, forkable flow. The paper is replaced by granular contributions (corrections, reviews, analyses, replications). The citation is replaced by a dependency graph: structural relationships, not flat counts. These metrics do not merely become less accurate; their unit of measurement ceases to exist. You cannot count ``papers'' when the unit of work is a tracked contribution, and you cannot count ``citations'' when the link between works is a structural dependency that specifies \emph{what} was used and \emph{how}, not a generic reference. A contribution graph, like the one software developers maintain, provides verified evidence of what someone actually did, a richer signal than a publication list, one that distinguishes between the tenth author on twenty papers and the person who made the key methodological correction that saved a research program. Contribution graphs can be gamed too, but gaming a tracked, public, auditable history is harder than gaming a self-reported CV.

This is the central disruption: the evaluation layer on which hiring, tenure, grants, and rankings are built assumes units that a visible flow dissolves. Adoption will require rethinking evaluation criteria, the same way open-access publishing required rethinking distribution models. A visible flow would also benefit funding decisions: a tracked record of contributions provides richer evidence of capability than a publication list, letting review panels focus on whether the proposed direction is worth pursuing.

% ============================================================================
% 5. Objections and Responses
% ============================================================================

\section{Objections and Responses}
\label{sec:objections}

\textbf{``Science is not software.''} Correct; the analogy is between workflows, not products. Science produces knowledge about the world; software produces tools that are installed and run. These are fundamentally different outputs. But the \emph{process} of producing them shares structural similarities: both involve complex, collaborative, error-prone intellectual work conducted over time. Open-source software has its own failures: critical vulnerabilities that persisted in widely used libraries, maintainer burnout, projects with zero external contributors. But even these failures illustrate the point. When serious bugs were discovered, the visible, tracked history allowed the community to trace exactly what happened, when the problem was introduced, and how to fix it. In science's current flow, an equivalent error would be a published paper with no way to trace the reasoning that produced it. Software's workflow is not perfect, but its failures are correctable precisely because the process is visible and tracked.

The deeper version of this objection concerns experimental science specifically. Bench work, fieldwork, and clinical research involve physical processes and tacit knowledge (buffer concentrations adjusted by feel, gel images interpreted by eye) that cannot be ``forked'' or ``diffed'' the way code can. Three responses. First, the objection proves too much: if some scientific knowledge cannot be recorded, the current system is even worse, because the published paper pretends to be a complete record while hiding far more than a tracked workflow would. A visible flow that captures 70\% of the process is still better than a paper that captures 5\%. Second, laboratory work is becoming increasingly digitized (electronic lab notebooks, automated instrumentation with logged parameters, computational image analysis), and the trend is toward the kind of digital process that can be tracked. Third, and most importantly, the paper is not arguing that every pipette movement must be recorded. There is enormous space between ``track every micro-decision'' and ``publish one polished narrative after two years of silence.'' A lab that documents its weekly decisions, failed protocols, revised hypotheses, and corrected analyses is already operating in a fundamentally different flow. The properties are continuous, not binary.

\textbf{``This will be gamed.''} Every metric is gamed. The question is whether a tracked, verified contribution history is harder to game than a line on a CV. It is, because the history is public and auditable.

\textbf{``Established institutions will resist.''} They always do, and the pattern is consistent across every major transition toward openness.

Open-source software faced the most extreme resistance; the argument was that volunteer-produced software could not compete with corporate teams. Today, Linux runs the majority of the world's servers and several of its former critics are among the largest open-source contributors. Open publishing followed: when arXiv launched in 1991, preprints were viewed with skepticism in many fields; today it hosts millions of papers and preprint posting is standard practice across physics, mathematics, and computer science. Open data followed: researchers would be scooped; today mandates are standard. Open code is the most recent wave and the most telling; the objections were ``my code is too messy to share'' and ``competitors will steal my methods''; today major ML conferences strongly encourage code submissions and journals increasingly mandate code availability. Each wave involved making \emph{process artifacts} more visible, and each time the prediction that openness would destroy quality was the opposite of what happened. Quality improved because visibility enabled scrutiny, correction, and broader contribution. Open process is the next step.

\textbf{``Process transparency will expose researchers.''} This objection reveals the defensive mindset the current flow produces. In a hidden-process system, exposure means punishment: someone finds your error, and the response is retraction, reputational damage, career risk. But in software, where the process has been visible for decades, the mindset is the opposite. Developers expect others to see their approach: how they addressed problems, what trade-offs they made, where they were uncertain. When someone forks their work to try a different interpretation of intermediate results, that is a compliment: it means the work was worth building on. The current system does not protect researchers; it conceals the process, so when errors are found, the response is maximally punitive. A visible process that shows an error was honest and was corrected is \emph{more} protective than a hidden one that, when exposed, looks like it might have been intentional.

\textbf{``I need to work privately while developing an idea.''} Of course. Software developers work in private repositories while exploring approaches and making mistakes, then make the repository public when ready, with the full history intact. The expectation is not that work happens in the open at every moment, but that when conclusions are shared, the full process that produced them is shared too. The infrastructure for this already exists and is trusted with far more commercially sensitive material than any research project. Commercial repository systems that host proprietary code for major technology companies provide granular access control that researchers can use identically. Moreover, many repositories never become fully public but are shared within a development community, a consortium, or an organization, and that is already a fundamentally different flow from one enclosed lab working in isolation. A research project visible to a multi-lab collaboration, a department, or a disciplinary community is not ``open to the world,'' but it is visible, trackable, and forkable within the community that matters. The properties are continuous, not binary.

\textbf{``Visible dead ends will discourage risk-taking.''} The concern is that if every failed approach is visible, researchers will avoid unconventional directions. The same concern was raised about open-source code, and the evidence went the other way: visibility normalized errors and encouraged bolder experimentation, because the community could see that dead ends were a routine part of every successful project. A visible record that shows a researcher tried five ambitious approaches, learned from each, and arrived at a sixth is more impressive, not less, than a polished paper that pretends the sixth approach was the only one considered.

% ============================================================================
% 6. Conclusion
% ============================================================================

\section{Conclusion}
\label{sec:conclusion}

The argument of this paper is not about platforms, tools, or infrastructure. It is about flow: the sequence of steps by which science is done, and the properties that sequence does or does not have.

Science's current flow is: work privately, publish a finished artifact, defend it, retract if wrong. This flow made sense when research was slow, teams were small, and publication was expensive. It does not make sense when AI generates plausible manuscripts faster than reviewers can check them, when research teams span continents, and when public trust depends on a process no one can see.

A different flow is possible: work visibly, track continuously, fork when directions diverge. Software development adopted this flow and the results are instructive: for the most widely used projects, quality improved as participation broadened. The same properties applied to science would accelerate progress, restore trust, scale quality assurance, and broaden participation; the mechanisms for each have been described in the preceding sections.

Millions of citizen scientists already contribute observations to research: hundreds of millions of biodiversity observations on iNaturalist, billions of bird sightings on eBird, millions of volunteers on Zooniverse. But the dominant model is data collection for someone else's study. A visible, trackable, forkable flow would bridge the gap between contributing data and contributing to research design, enabling collaborative science at a scale the current workflow cannot support.

The transition does not require abandoning existing institutions; it requires that the option exist. Adoption pressure will follow, as it did for open-access publishing and open-source software.

Science opened its outputs. The next step is to open its process.

% ============================================================================
% References
% ============================================================================

\bigskip
\noindent\textbf{Note on references.} Every reference in this paper is available in full text without institutional access or payment. While other influential work exists on relevant topics, we deliberately include only sources that any reader can obtain and verify.

\bibliographystyle{plainnat}
\bibliography{paper_opening_science_process}

\end{document}